\documentclass[preprint, showpacs,preprintnumbers,amsmath,amssymb,nofootinbib]{revtex4}
\usepackage{amssymb}
\usepackage{amsfonts}

 \usepackage{epsf}
 \usepackage{graphicx}
\usepackage{float}
\begin{document}
\newcommand{\be}[1]{\begin{equation}\label{#1}}
 \newcommand{\ee}{\end{equation}}
 \newcommand{\bea}{\begin{eqnarray}}
 \newcommand{\eea}{\end{eqnarray}}
 \newcommand{\bed}{\begin{displaymath}}
 \newcommand{\eed}{\end{displaymath}}
 \def\disp{\displaystyle}

 \def\gsim{ \lower .75ex \hbox{$\sim$} \llap{\raise .27ex \hbox{$>$}} }
 \def\lsim{ \lower .75ex \hbox{$\sim$} \llap{\raise .27ex \hbox{$<$}} }

\title{Cosmic opacity: cosmological-model-independent tests and their impacts on cosmic acceleration}

\author{Zhengxiang Li$^{1,3}$, Puxun Wu$^{2}$,  Hongwei Yu$^{1, 2,}$\footnote{Corresponding author: hwyu@hunnu.edu.cn} and Zong-Hong Zhu${^3}$ }

\address{$^1$Department of Physics and Key Laboratory of Low
Dimensional Quantum Structures and Quantum Control of Ministry of
Education, \\Hunan Normal University, Changsha, Hunan 410081, China
\\$^2$Center of Nonlinear Science and Department of Physics, Ningbo
University,  Ningbo, Zhejiang 315211, China
\\$^3$ Department of Astronomy, Beijing Normal University, Beijing 100875, China}

\begin{abstract}
With assumptions that the violation of the distance-duality relation
entirely arises from  non-conservation of the photon number and the
absorption is frequency independent in the observed frequency range,
we perform cosmological-model-independent tests for the cosmic
opacity. The observational data include the largest Union2.1 type Ia
supernova sample, which is taken  for observed $D_\mathrm{L}$, and
galaxy cluster samples compiled by De Filippis {\it et al.} and
Bonamente {\it et al.}, which are responsible for providing observed
$D_\mathrm{A}$. Two parameterizations, $\tau(z)=2\epsilon z$ and
$\tau(z)=(1+z)^{2\epsilon}-1$ are adopted for the optical depth
associated to the cosmic absorption. We find that, an almost
transparent universe is favored by Filippis {\it et al.} sample but
it is only marginally accommodated by Bonomente {\it et al.} samples
at 95.4\% confidence level (C. L.) (even at 99.7\% C. L. when the
$r<100~ \mathrm{kpc}$-cut spherical $\beta$ model is considered).
Taking the possible  cosmic absorption (in 68.3\% C. L. range)
 constrained from the model-independent tests
into consideration, we correct the distance modulus of SNe Ia and
then use them to study their cosmological implications. The
constraints on the $\Lambda$CDM show that a decelerating expanding
universe with $\Omega_\Lambda=0$ is only allowed at 99.7\% C. L. by
observations when the Bonamente {\it et al.} sample is considered.
Therefore, our analysis suggests  that an accelerated cosmic
expansion is still needed to account for the dimming of SNe and the
standard cosmological scenario remains being supported by current
observations.

\end{abstract}


 \maketitle
 \renewcommand{\baselinestretch}{1.5}

\section{INTRODUCTION}
The type Ia supernovae (SNe Ia) are observed to  be fainter than
expected from the luminosity-redshift relationship in a decelerating
universe. This unanticipated dimming was first attributed to an
accelerating expansion of the universe~\cite{Riess0, Perlmutter}.
Although the existence of cosmic acceleration has been verified by
several other observations, initially, there had been some debates
on the interpretation of underlying physical mechanism for the
observed SNe Ia dimming. For example, dust in the Milk Way and
oscillation of photons propagating in extragalactic magnetic fields
into very light axions had  been proposed to account for the
dimming~\cite{Aguirre, Csaki}. Any kind of photon number violation,
such as absorption, scattering or axion-photon mixing, sensibly
imprints its influence on the Tolman test~\cite{Tolman}, which can
be rewritten as a relationship among cosmological distance
measurements known as the famous distance-duality (DD)
relation~\cite{Etherington1,Etherington2,Ellis},
\begin{equation}
\frac{D_\mathrm{L}}{D_\mathrm{A}}(1+z)^{-2}=1,
\end{equation}
where $z$ is the redshift, $D_\mathrm{L}$ and $D_\mathrm{A}$ are the
luminosity distance and the angular diameter distance (ADD)
respectively. This reciprocity relation holds for general metric
theories of gravity in any background, not just in that of the
Friedmann-Lem$\hat{\mathrm{a}}$itre-Robertson-Walker background, and
it is also valid for all cosmological models based on the Riemannian
geometry. That is, its validity depends neither on Einstein field
equations nor on the nature of the matter-energy content. The DD
relation plays an important role in modern
cosmology~\cite{Schneider,Cunha,Mantz,Komatsu}, and, in most cases,
it has  been applied, without any doubt, to analyze the cosmological
observations. However, the reciprocity relation may be violated if
photons do not travel on null geodesics or the universe is opaque.

Fortunately, it is in principle possible to perform a valid check on
the DD relation by means of astronomical observations. The basic
idea is to search for observational candidates with known intrinsic
luminosities as well as intrinsic sizes, and then determine their
$D_\mathrm{L}$ and $D_\mathrm{A}$ to test the Etherington relation
directly. It is difficult for us to find objects of the same class
with both known intrinsic luminosities and intrinsic sizes. Thus,  a
$\Lambda$CDM cosmological model is usually assumed when one performs
tests by utilizing observed $D_\mathrm{L}$ or
$D_\mathrm{A}$~\citep{Bassett1,Bassett2,Uzan,De bernardis} and the
results show that there is no strong evidence for deviations from
the standard DD relation. Recently, a model-independent method has
been proposed to test DD relation by considering two different
classes of objects, for example, SNe Ia and galaxy clusters, from
which $D_\mathrm{L}$ and $D_\mathrm{A}$ are determined
separately~\cite{Holanda1,Holanda2,Zhengxiang,Nair}. For a given ADD
data, in order to obtain the corresponding $D_\mathrm{L}$ from SNe
Ia, a selection criteria $\Delta z=
\left|z_{\mathrm{Cluster}}-z_{\mathrm{SNe~Ia}}\right|\leq0.005$ is
adopted. Using the phenomenological parameterized forms
\begin{equation}
\frac{D_\mathrm{L}}{D_\mathrm{A}}(1+z)^{-2}=\eta(z)\;,
\end{equation}
with $\eta(z)=1+\eta_0z$ and $\eta(z)=1+\frac{z}{1+z}\eta_0$, and
the data from Union2 SN Ia~\cite{Amanullah} and galaxy clusters, it
was found that the DD relation can be accommodated at $1\sigma$ C.
L. for the elliptical $\beta$ model~\cite{Filippis} and at $3\sigma$
C. L. for the spherical $\beta$ model~\cite{Bonamente}. More
recently, in order to avoid the bias from the redshift difference
between $D_\mathrm{L}$ and ADD, methods, such as, the binning of SNe
Ia~\cite{Hu}, the interpolation~\cite{Liang} and local
regression~\cite{Cardone} from nearby SNe Ia points for a given
galaxy cluster, were proposed and similar results were obtained. In
addition, The DD relation tests by use of SNe Ia and gas mass
function of galaxy clusters were carried out and similar results
were also obtained~\cite{Holandad, Holandac, Goncalves}. So,
overall, all the tests performed so far show that there is no strong
indication of the DD relation violation. However, let us note that
systematic uncertainties resulting from the morphological models of
galaxy clusters and the redshift difference between $D_\mathrm{L}$
and ADD might exert influences on DD relation tests.

If one considers that the photon traveling along null geodesic is
unassailable, the DD relation violation most likely implies
non-conservation of the photon number which has a mundane origin
(scattering from dust or free electron) or an exotic origin (photon
decay or photon mixing with other light states such as the dark
energy, dilaton or axion~\cite{Csaki,Bassett2}). In this case, the
flux received by the observer will be reduced and so the universe is
opaque. If we assume that the flux from the source is decreased by a
factor $e^{-\tau(z)}$, then  the inferred (observed) luminosity
distance  differs from the ``true" one~\cite{More, Chen1, Chen2}
\begin{equation}{\label{eq3}}
D_{\mathrm{L,obs}}=D_{\mathrm{L,true}}\cdot e^{\tau/2}\;,
\end{equation}
where $\tau$ is the opacity parameter which denotes the optical
depth associated to the cosmic absorption. Initially, More {\it et
al.}~\cite{More} studied the cosmic opacity by examining the
difference of the opacity parameter at redshifts $z=0.20$ and
$z=0.35$, $\Delta\tau=\tau(0.35)-\tau(0.2)$, where the difference of
the observational luminosity distance ($\Delta D_{\mathrm{L,obs}}$)
at these two redshifts was estimated from two subsamples of ESSENCE
SN Ia~\cite{Davis} and the corresponding $\Delta
D_{\mathrm{L,true}}$ was derived from the distance measurements of
baryonic acoustic feature~\cite{Percival} in the context of
$\Lambda$CDM. Assuming flat priors on $\Omega_\Lambda$ and
$\Omega_\mathrm{M}$ in the range of $0<\Omega<1$, and uniformly
spaced values of $\Delta\tau\in[0,0.5]$, they found that a
transparent universe is favored (posterior probabilities of
$\Delta\tau$ peaked at 0) and $\Delta\tau<0.13$ at 95\% C. L.. This
method has been applied to investigate the homogeneity of the cosmic
opacity in different redshift regions and the results suggest that
the cosmic opacity oscillates between zero and non-zero values as
redshift varies~\cite{Jun,Nair2}. Later, Avgoustidis {\it et
al.}~\cite{Avgoustidis1,Avgoustidis2} carried out further studies by
assuming an optical depth parameterization $\tau(z)=2\epsilon z$ or
$\tau(z)=(1+z)^{2\epsilon}-1$ for small $\epsilon$ and $z\leq1$.
There they took the standard luminosity distance in the spatially
flat $\Lambda$CDM ($(1+z)^2D_\mathrm{A}(z,\Omega_\mathrm{M})$) and
the Union SN Ia~\cite{Kowalski} for $D_{\mathrm{L,true}}$ and
$D_{\mathrm{L,obs}}$ respectively. In addition to the SNe Ia data,
they also used the measurements of the cosmic expansion
$H(z)$~\cite{Stern, Riess1}. By taking $\epsilon\in[-0.5,0.5]$,
$\Omega_\mathrm{M}\in[0,1]$ and $H_0\in[74.2-3\times3.36,
74.2+3\times3.36]$~\cite{Riess1} all uniformly spaced over the
relevant intervals in a flat $\Lambda$CDM model and performing a
full Bayesian likelihood analysis, they obtained a result
$\epsilon=-0.04^{+0.08}_{-0.07}$ ($2\sigma$ C. L.), which
corresponds to an opacity $\Delta\tau<0.012$ (95\% C. L.) for the
redshift range between 0.2 and 0.35, almost a factor of 11 stronger
than the constraint obtained in Ref.~\cite{More}. Recently, Lima
{\it et al.}~\cite{Lima} reexamined this issue by confronting  the
luminosity distance which is dependent on two free parameters, i.e.,
the so called cosmic absorption parameter ($\alpha_*$) and the
matter density ($\Omega_\mathrm{M}$), with observations, using a
subsample of Union2 SN Ia obtained by selecting SNe Ia with
redshifts greater than $cz = 7000km/s$ in order to avoid effects
from Hubble bubble, and they found that the Einstein-de Sitter model
($\Omega_\mathrm{M}=1$) could be allowed at 68.3\% (95.4\%) C. L. in
the case of a constant (epoch-dependent) absorption and concluded
that a cosmic absorption may be responsible for the dimming of the
distant SNe Ia without the need of an accelerated expansion of the
universe. However, all these studies concerning the cosmic opacity
assume a (flat) $\Lambda$CDM model, and are thus model-dependent.

Here we propose another model-independent method to examine the
cosmic opacity and investigate its possible implications for the
cosmic evolution. If we assume that the violation of the DD relation
is purely caused by the photon number non-conservation, then we can
find out whether the universe is opaque by checking  the possible
violation of the DD relation. It should be emphasized that the
cosmic absorption not only affects the luminosity distance
measurements of SNe Ia observations as shown in Eq.~(\ref{eq3}), but
also exerts influences on the angular diameter distance measurements
determined from SZE+X-ray surface brightness
observations~\cite{Sunyaev,Cavaliere},
\begin{equation}
D_\mathrm{A}\propto\frac{\Delta T^2_{\mathrm{CMB}}}{S_\mathrm{X}},
\end{equation}
where $\Delta T_{\mathrm{CMB}}$ is the temperature change due to the
Sunyaev-Zel'dovich effect (SZE) when the cosmic microwave background
(CMB) photons pass through the hot intracluster medium and
$S_\mathrm{X}$ is the X-ray surface brightness of galaxy clusters.
The SZE spectra distortion of the CMB is determined by measuring the
intensity decrements, $\Delta I$, which is sensitive to the cosmic
absorption. Additionally, the surveys of X-ray surface brightness
are also sensitive to the opacity of the universe. Supposing the
absorption is frequency independent in the observed frequency range
(from microwave band to X-ray band), the ``true" ADD connects the
observed one measured in an opaque universe with
$D_{\mathrm{A,true}}=D_\mathrm{A}^{\mathrm{cluster}}\cdot e^{\tau}$.
Thus, in actual calculations, the DD relation takes the following
form:
\begin{equation}
\frac{D_\mathrm{L}^{\mathrm{SN}}}{D_\mathrm{A}^{\mathrm{cluster}}}(1+z)^{-2}=e^{3\tau/2}.
\end{equation}
Now we will use the the largest Union2.1 SN Ia sample~\footnote{See
Ref.~\cite{SCP}}~\cite{Suzuki} and the ADD data from galaxy cluster
samples~\cite{Filippis,Bonamente} to test, model-independently, the
possible violation  of the DD relation, which can be translated to a
possible cosmic opacity. The observed $D_\mathrm{L}$ and
$D_\mathrm{A}$ come  from the latest Union2.1 SN Ia and galaxy
clusters samples~\cite{Filippis,Bonamente}, respectively. Actually,
there are a number of inherent uncertainties in the selected
astrophysical objects from which the observed $D_\mathrm{A}$ are
derived, e. g., the cluster asphericity~\cite{Filippis} and the
model for the cluster gas distribution~\cite{Bonamente}. In this
paper, we consider the elliptical $\beta$ model galaxy clusters
sample~\cite{Filippis}, spherical $\beta$ model,
$r<100~\mathrm{kpc}$-cut spherical $\beta$ model and hydrostatic
equilibrium model galaxy clusters samples~\cite{Bonamente} to
investigate the impact of these inherent uncertainties on the cosmic
opacity test.

\section{DATA AND CONSTRAINT RESULTS}
\begin{figure*}
\centering
\includegraphics[width=0.40\linewidth]{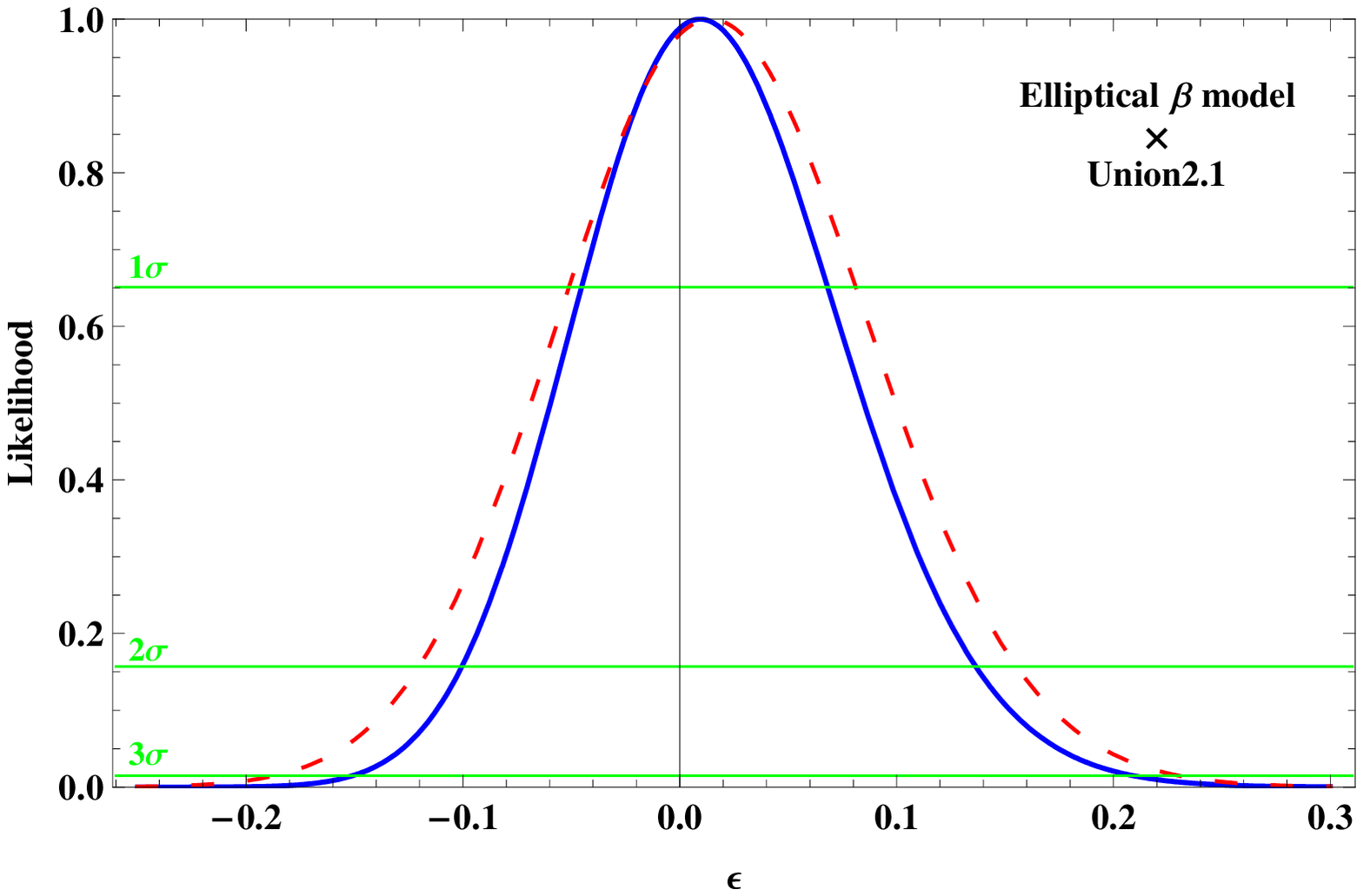}
\caption{ Probability distribution functions of opacity parameter
$\epsilon$ obtained from the De Filippis {\it et al.} sample and
Union2.1 SN Ia pairs for two parameterizations: $\tau(z)=2\epsilon
z$ (blue solid) and $\tau(z)=(1+z)^{2\epsilon}-1$ (red dashing).}
\label{Fig1}
\end{figure*}

\begin{figure*}
   \centering
       \includegraphics[width=0.32\linewidth]{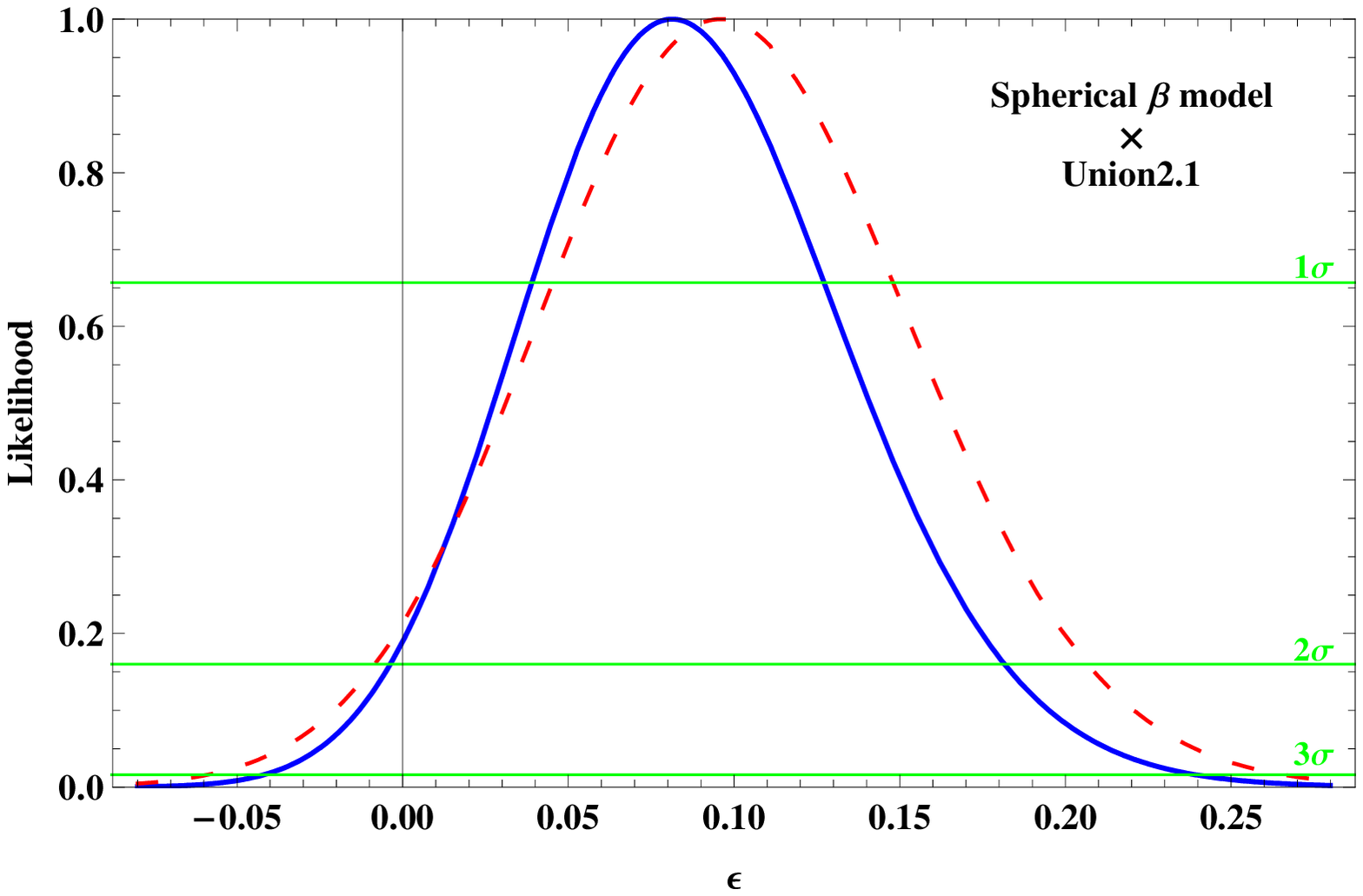}
       \includegraphics[width=0.32\linewidth]{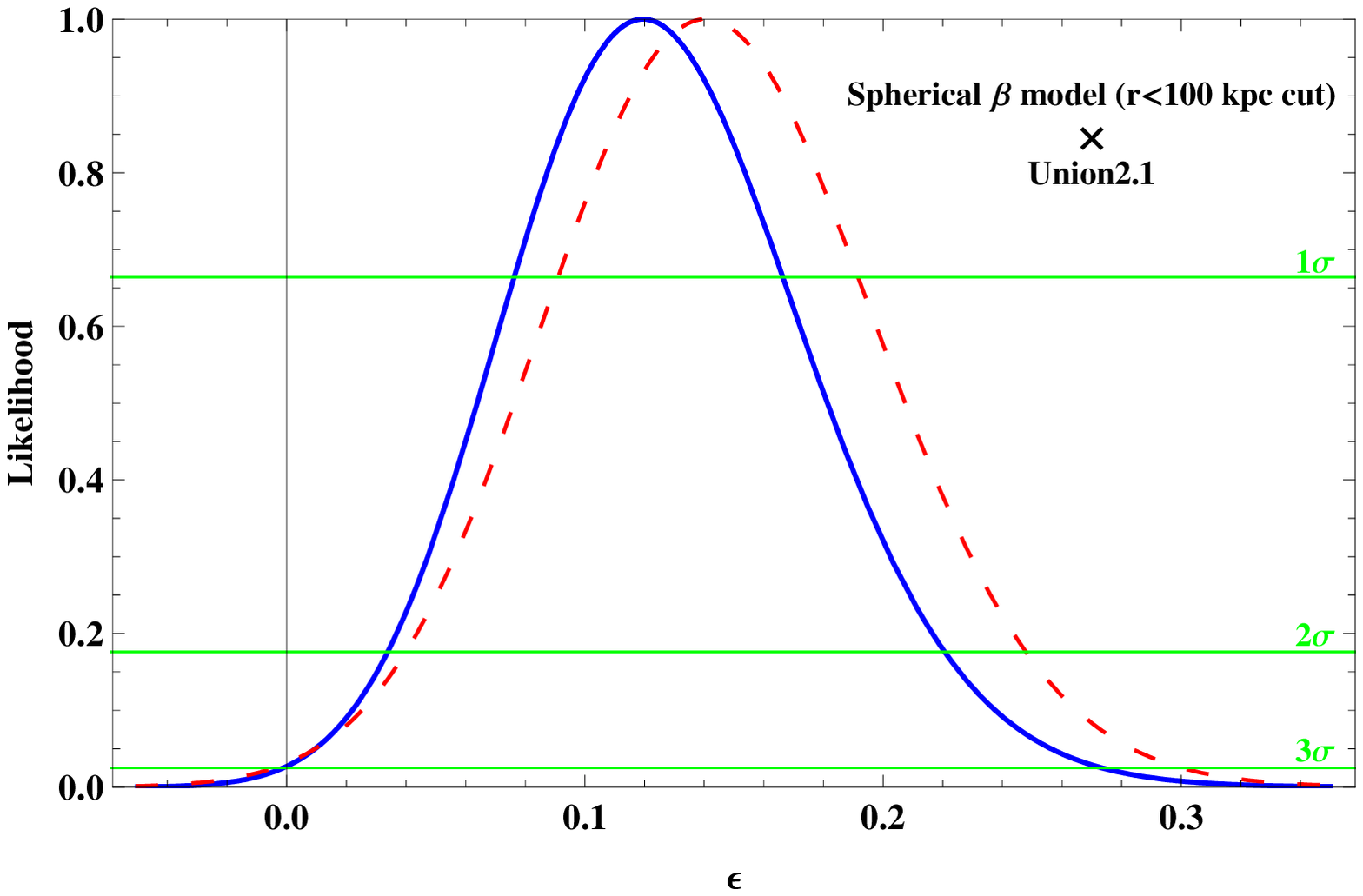}
       \includegraphics[width=0.32\linewidth]{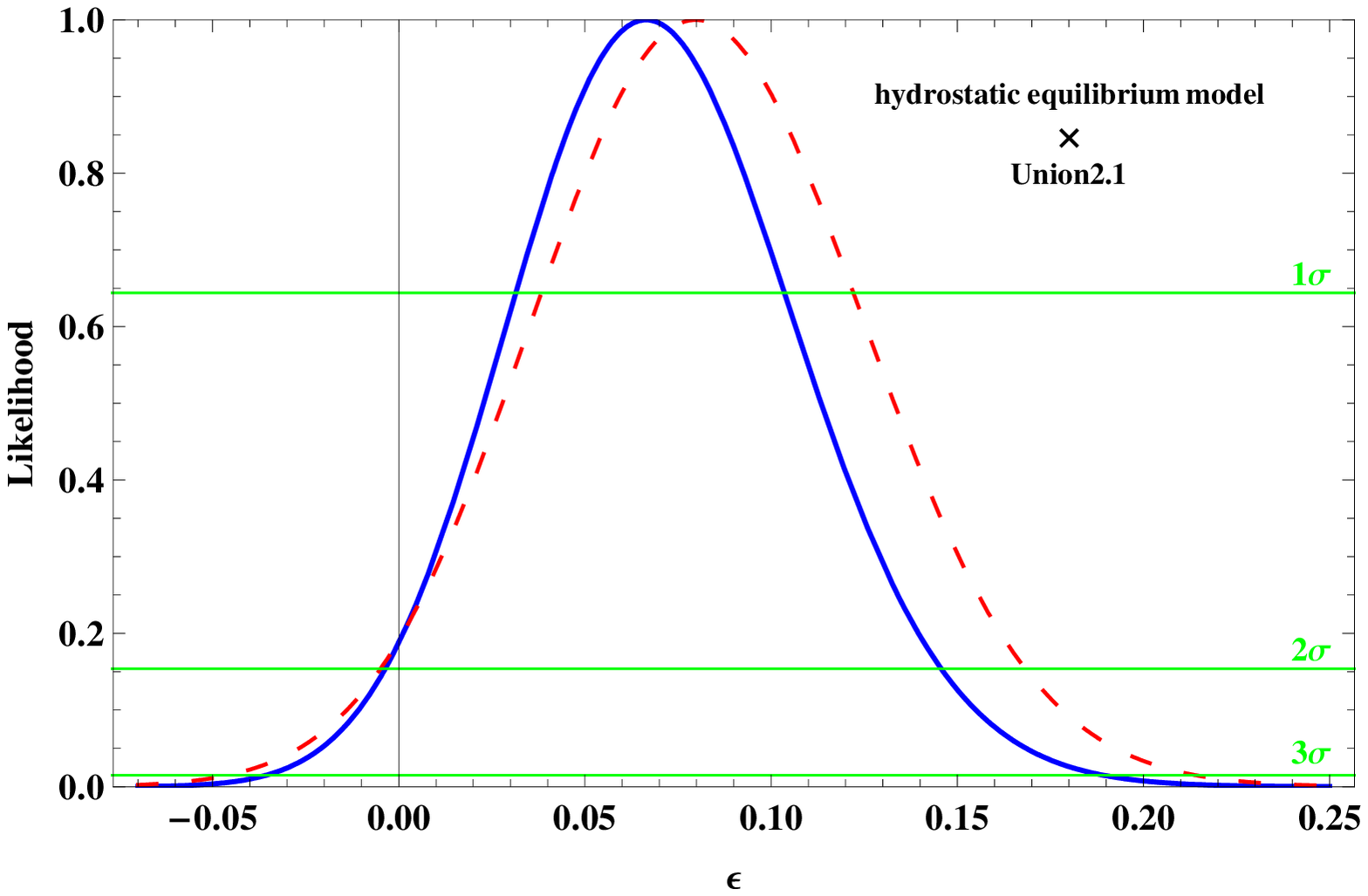}
   \caption{ Probability distribution functions of opacity parameter $\epsilon$ obtained from the Bonamente {\it et
al.} samples and Union2.1 SN Ia pairs for two parameterizations:
$\tau(z)=2\epsilon z$ (blue solid) and $\tau(z)= (1 +
z)^{2\epsilon}-1$ (red dashing). The left, middle, and right panel
represent results constrained from the isothermal $\beta$, $r<100$
kpc-cut isothermal $\beta$ and hydrostatic equilibrium model
respectively. } \label{Fig2}
\end{figure*}

In order to place constraints on the cosmic opacity parameter
$\tau$, we first parameterize it with two monotonically increasing
functions of redshift, i.e., $\tau(z)=2\epsilon z$ and
$\tau(z)=(1+z)^{2\epsilon}-1$~\cite{Avgoustidis1}. These two
parameterizations are basically similar for $z\ll1$ but they differ
when $z$ is not very small. As the data applied in our following
analysis discretely distribute in the redshift range $0.023\leq
z\leq0.890$, our analysis we are going to perform may tell us
something about the possible dependence of  the test results on the
parametric forms for $\tau$. To obtain $\tau_{\mathrm{obs}}(z)$
determined by the following expression:
\begin{equation}
\tau_{\mathrm{obs}}(z)=\frac{2}{3}\ln
\bigg[\frac{D^{\mathrm{SN}}_\mathrm{L}}{D^{\mathrm{cluster}}_\mathrm{A}(1+z)^2}\bigg],
\end{equation}
the data pairs of observed $D_\mathrm{L}$ and $D_\mathrm{A}$ almost
at the same redshift should be supplied. For the observed
$D_\mathrm{L}$, the largest Union2.1 SN Ia is considered. Galaxy
cluster samples, where the $D_\mathrm{A}$ are obtained by combining
the SZE+X-ray surface brightness
measurements~\cite{Sunyaev,Cavaliere}, are responsible for providing
the observed $D_\mathrm{A}$. The first one, including a selection of
7 clusters compiled by Mason {\it et al.}~\cite{Mason} and a sample
of 18 clusters collected by Reese {\it et al.}~\cite{Reese}, was
re-analyzed by Filippis {\it et al.}~\cite{Filippis} by assuming an
elliptical $\beta$ model for the galaxy clusters. The second kind of
samples are compiled by Bonamente {\it et al.}~\cite{Bonamente} with
three different models for the cluster plasma and dark matter
distribution, i.e., the spherical $\beta$ model, spherical $\beta$
model with $r<100$ kpc cut, and hydrostatic equilibrium model.
Therefore, the data derived from these three different models are
adopted to check whether the cosmic opacity tests are sensitive to
the model for the cluster gas distribution. The observed
$D_\mathrm{L}$ are binned from the data points of Union2.1 SN Ia
with their redshifts satisfying the certain criteria ($\Delta
z_{\mathrm{max}}=
\left|z_{\mathrm{cluster}}-z_{\mathrm{SNe~Ia}}\right|_{\mathrm{max}}\leq0.005$)
to match the observational data of the ADD
samples~\cite{Bevington,Hu}. This binning method can minimize the
statistical errors originating from the redshift difference between
$D_\mathrm{L}$ and $D_\mathrm{A}$. On the other hand, we alter
$\Delta z_{\mathrm{max}}$ from 0.000 to 0.005 to ensure the number
of the clusters that share the same SNe to be as few as possible, so
as to reduce the dependence of opacity tests on the correlation of
redshift-matched SNe. After obtaining $\tau_{\mathrm{obs}}(z)$ from
these selected data pairs, we estimate the free parameters of a
given parametric form by using the standard minimum $\chi^2$ route:
\begin{equation}
\chi^2(z; \mathbf{p})=\sum_i \frac{[\tau(z;
\mathbf{p})-\tau_{\mathrm{obs}}(z)]^2}{\sigma^2_{\tau_{\mathrm{obs}}}},
\end{equation}
where $\sigma_{\tau_{\mathrm{obs}}}$ is the error of
$\tau_{\mathrm{obs}}$ associated with the observed $D_\mathrm{L}$
and $D_\mathrm{A}$, and $\mathbf{p}$ represents the free parameters
to be constrained. The graphic representations and numerical results
of the probability distribution of the opacity parameter $\epsilon$
constrained from the model-independent tests are shown in
Figures~(\ref{Fig1},~\ref{Fig2}) and Table~(\ref{Tab1}). These
suggest that the dependence of test results on the above-chosen
parameterizations for $\tau(z)$ is relatively weak. Similar to the
results obtained by examining the cosmic opacity in a particular
redshift range (0.20-0.35)~\cite{More,Avgoustidis1,Avgoustidis2} and
deforming the DD relation in terms of the cosmic absorption
parameter~\cite{Lima}, we find, from Figure (\ref{Fig1}), that an
almost transparent universe is also favored by the elliptical
$\beta$ model galaxy clusters sample~\cite{Filippis}. For the ADD
samples given by Bonamente {\it et al.}~\cite{Bonamente}, the
results are shown in Figure~\ref{Fig2}. We find that the results are
nearly not sensitive to the model of cluster gas distribution and a
transparent universe can only be marginally accommodated at 95.4\%
C. L. (even at 99.7\% C. L. when the $r<100$ kpc-cut spherical
$\beta$ model is considered). That is, all the constraints on the
opacity parameter obtained from the Bonamente {\it et al.} samples
prefer an opaque universe. These results are clearly different from
what were obtained based on the $\Lambda$CDM in
Refs.~\cite{More,Lima}, where a transparent universe is obviously
supported. In fact, these cosmic opacity test results are very
similar to the previous model-independent tests for the DD
relation~\cite{Holanda1,Holanda2,Zhengxiang,Hu,Liang}. However, our
objective here is the cosmic opacity test with an assumption that
the violation of the DD relation entirely originates from the
non-conservation of photon number, rather than the DD relation test
itself.

\begin{figure*}
   \centering
   \includegraphics[width=0.8\linewidth]{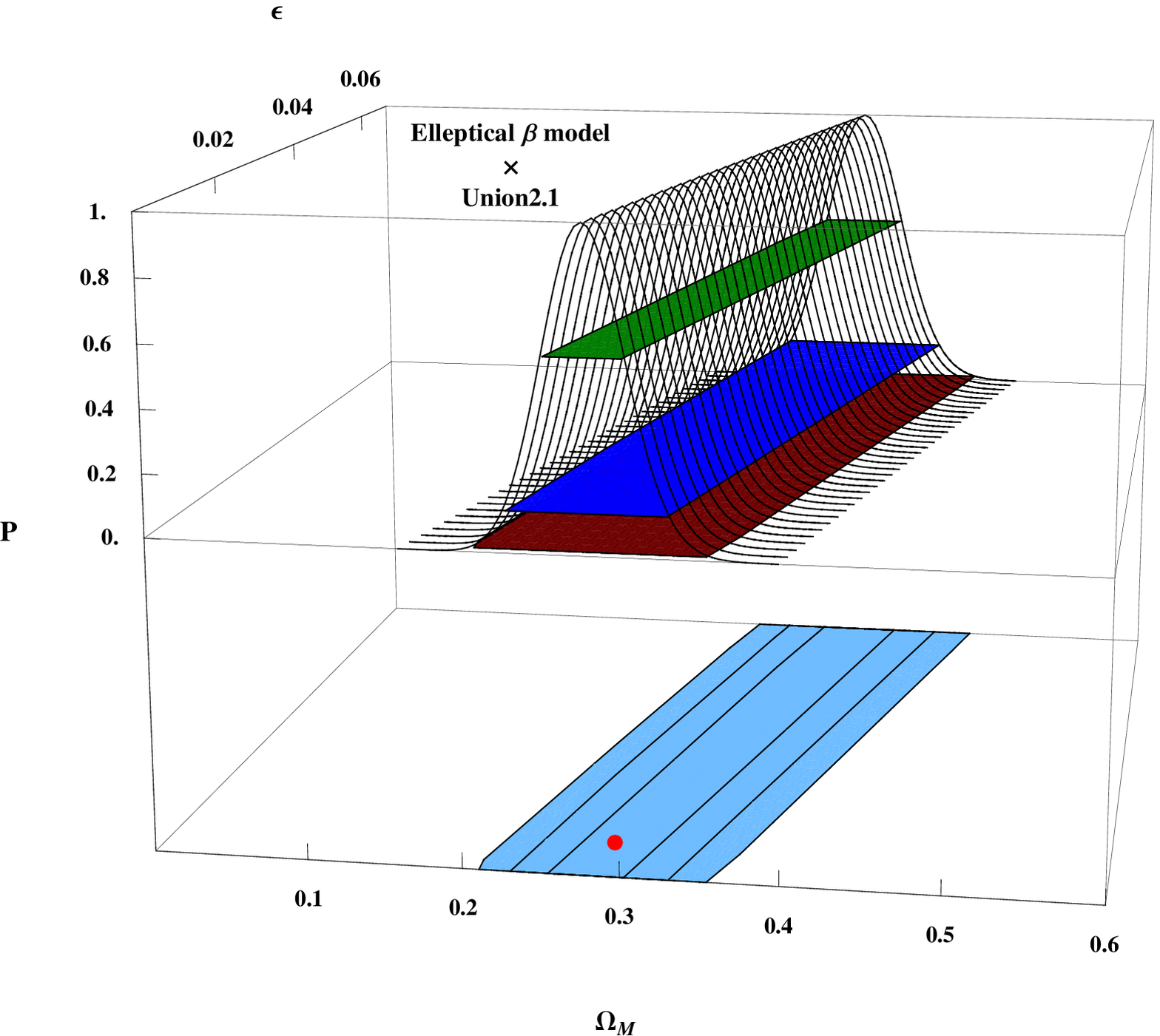}
   \caption{ {\bf Upper:} The probability distributions of $\Omega_\mathrm{M}$ in the context of flat $\Lambda$CDM when
   the absorptions model-independently constrained from
   the combination of De Filippis {\it et al.} sample and Union2.1 SN
   Ia are considered. The green, blue and red zonal regions represent the spans of $\Omega_\mathrm{M}$
   at 68.3\%, 95.4\% and 99.7\% C. L. respectively. As the results are not sensitive to the parametric
   form of $\tau(z)$, here the linear expression ($\tau(z)=2\epsilon z$) is
   applied. {\bf Lower:} The projections of the upper zonal regions in the $\Omega_\mathrm{M}-\epsilon$ plane. The
   red dot ($\Omega_\mathrm{M}$=0.285, $\epsilon$=0.009) represents the
    best fit case.} \label{Fig3}
\end{figure*}

\begin{figure*}
\centering
\includegraphics[width=0.8\linewidth]{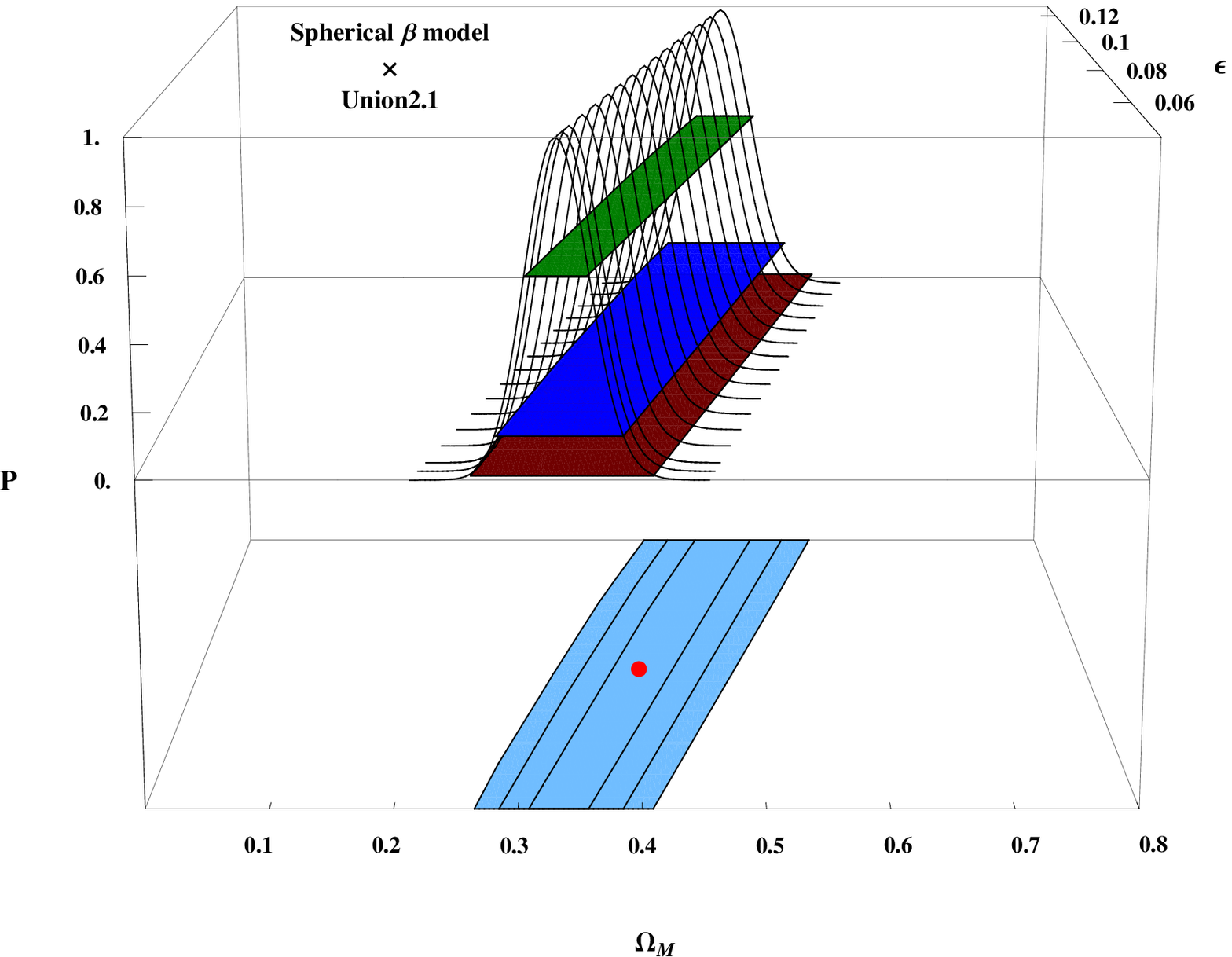}
   \caption{{\bf Upper:} The probability distributions of $\Omega_\mathrm{M}$ in the context of flat $\Lambda$CDM when
   the absorptions model-independently constrained from
   the combination of Bonemente {\it et al.} sample and Union2.1 SN
   Ia are considered. The green, blue and red zonal regions represent the spans of $\Omega_\mathrm{M}$
   at 68.3\%, 95.4\% and 99.7\% C. L. respectively. As the results are not sensitive to the parametric
   form of $\tau(z)$, here the linear expression ($\tau(z)=2\epsilon z$) is
   applied. {\bf Lower:} The projections of the upper zonal regions in the $\Omega_\mathrm{M}-\epsilon$ plane. The
   red dot ($\Omega_\mathrm{M}$=0.397, $\epsilon$=0.081) represents the
   best fit case.} \label{Fig4}
\end{figure*}

In order to explore the implications of the cosmic opacity, let us
transform the SNe Ia distance modulus in a transparent universe into
that in an opaque one
\begin{equation}
\mu_{\mathrm{true}}(z)=\mu_\mathrm{obs}(z)-2.5[\log e]\tau(z)\;,
\end{equation}
and study the  cosmological constraints resulting from this
correction. Since the high redshift galaxy cluster data are absent
in our discussion of the cosmic opacity, the
distance-modulus-modified SNe Ia used to investigate the
cosmological constraints are cut down from Union2.1 with the
criteria $z\leq0.784$ and $z\leq0.890$ when clusters in the Filippis
{\it et al.} sample and the Bonamente {\it et al.} samples are
applied, respectively. Since the dependence of cosmic opacity tests
on the above-chosen parametric forms of $\tau(z)$ is relatively
weak, we only consider results obtained from the linear
parametrization ($\tau(z)=2\epsilon z$) in the following
cosmological implication analysis. For the flat $\Lambda$CDM model,
different from the methods used in
Refs.~\cite{Avgoustidis1,Avgoustidis2,Lima}, we examine the
probability distributions of $\Omega_\mathrm{M}$ by considering the
possible (68.3\% C. L. range) opacity parameter $\epsilon$
constrained from the previous model-independent tests. The results
are shown in Figures~(\ref{Fig3}, \ref{Fig4}). We find that the
opacity parameter $\epsilon$ constrained from previous
cosmological-model-independent tests impact slightly on the
likelihood distributions of $\Omega_\mathrm{M}$ and a universe with
$\Omega_\Lambda>0$ is required to account for the dimming of the SNe
Ia. This differs from the results in Ref.~\cite{Lima}, where the
Einstein-de Sitter universe ($\Omega_\mathrm{M}=1$) can be easily
accommodated at 68.3\% and 95.4\% C. L. for the constant and
epoch-dependent absorptions, respectively.

\begin{figure*}
   \centering
       \includegraphics[width=0.40\textwidth]{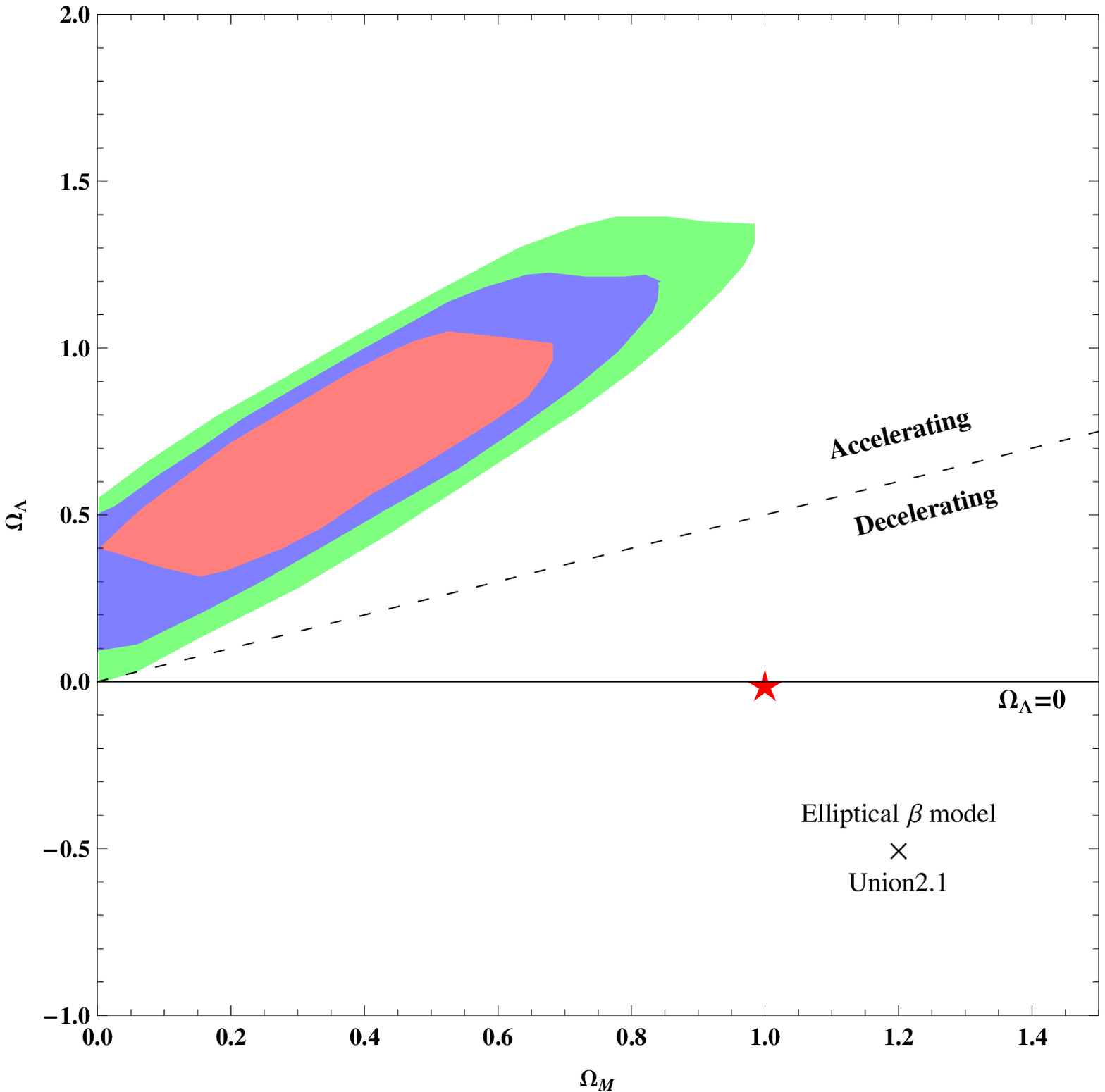}
       \includegraphics[width=0.40\textwidth]{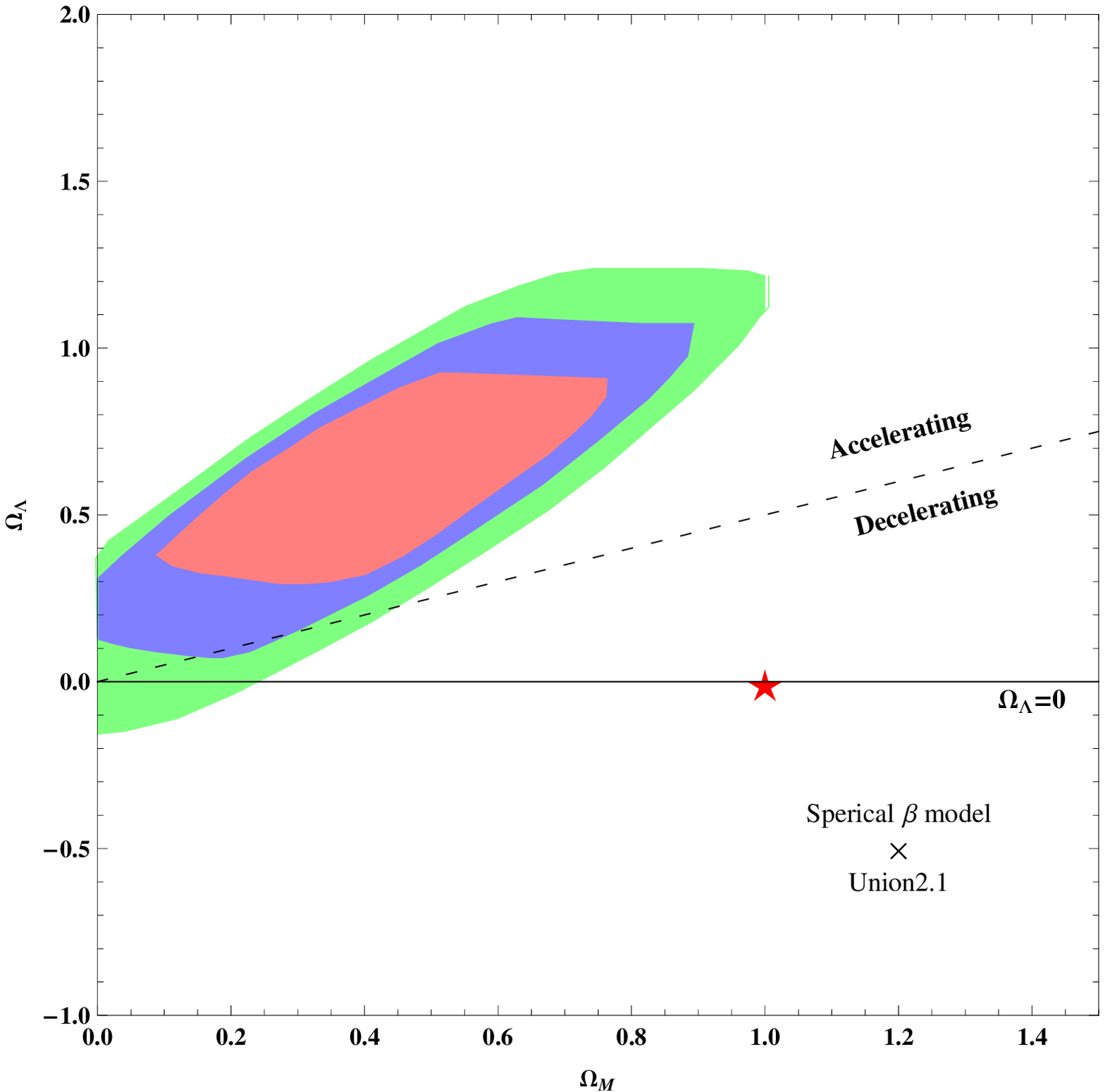}
   \caption{ Marginalized regions at 68.3\%, 95.4\% and 99.7\% C. L. in the $\Omega_\mathrm{M}-\Omega_\Lambda$ plane for the corrected data of subsamples of
    Union2.1 SN Ia by considering
    the observational constrained cosmic absorptions. As the results are not sensitive to the parametric form of
   $\tau(z)$, here the linear expression ($\tau(z)=2\epsilon z$) is
   applied. The red stars ($\Omega_\mathrm{M}=1.0$, $\Omega_\Lambda=0.0$) represent the Einstein-de Sitter universe. The left(right)
    panel corresponds to the results obtained from the De Filippis {\it et al.}( Bonamente {\it et al.})
    sample.} \label{Fig5}
\end{figure*}

Without a spatially flat universe prior, we also investigate the
$\Lambda$CDM with the corrected distance modulus of Union2.1 SN Ia
by taking the opacity parameter $\epsilon$ in 68.3\% C. L. range
constrained from the previous model-independent tests into
consideration. The linear parametrization for the cosmic opacity is
also considered. The results projecting to the
$\Omega_\mathrm{M}-\Omega_\Lambda$ plane are shown in
Figure~\ref{Fig5}. Because of the ``marginalization" of $\epsilon$,
which somewhat weakens the constraints on parameters
$\Omega_\mathrm{M}$ and $\Omega_\Lambda$, the statement that the
expansion of universe is accelerating is less eloquent than the one
in Ref.~\cite{Riess0}, which concludes that a currently accelerating
universe is needed at 99.9\% (3.9$\sigma$) C. L. to agree with their
distance measurements of SNe Ia. However, we find that a
decelerating universe with $\Omega_\Lambda=0$ is only allowed at
$99.7\% $ C. L. by observations when the spherical $\beta$ model is
taken into account. So, the standard cosmological scenario is still
supported by observations, although current data may favor a
universe with non-zero opacity,

\section{CONCLUSION AND DISCUSSION}
In this paper, by considering the luminosity distances provided by
the largest Union2.1 SN Ia sample together with the ADD given by
galaxy clusters samples, we first examine the possible cosmic
opacity in a cosmological-model-independent way. Two
redshift-dependent parametric expressions: $\tau(z)=2\epsilon z$ and
$\tau(z)=(1+z)^{2\epsilon}-1$ are considered to describe the optical
depth associated to the cosmic absorption. The results suggest that
the tests of cosmic opacity are not significantly sensitive to the
parametrization for $\tau(z)$. For the ADD sample compiled by
Filippis {\it et al.}~\cite{Filippis} with an elliptical $\beta$
model, we obtain that a universe with little opacity (almost
transparent) is favored. For the ADD samples given by Bonamente {\it
et al.}~\cite{Bonamente}, where three different cluster gas
distribution models are applied, the test results suggest that a
transparent universe can only be marginally consistent with
observations at 95.4\% C. L. (even at 99.7\% C. L. as the $r<100$
kpc-cut spherical $\beta$ model is applied). These are fairly
different from the conclusions in Refs.~\cite{More,Lima}. By
considering the possible cosmic opacity (68.3\% C. L. range)
constrained from the previous model-independent tests, we obtain the
corrected distance modulus of SNe Ia and then use them to
investigate its cosmological implications. In the context of a flat
$\Lambda$CDM, the likelihood functions of $\Omega_\mathrm{M}$ are
examined. The results are shown in Figures~(\ref{Fig3},~\ref{Fig4}).
We find that the opacity parameter $\epsilon$ constrained from
previous cosmological-model-independent tests have slight influences
on the probability distributions of $\Omega_\mathrm{M}$ and a
universe with $\Omega_\Lambda>0$ is required to account for the
dimming of the SNe Ia. Discarding the condition of the spatial
flatness, we display the corresponding plots in the
$\Omega_\mathrm{M}-\Omega_\Lambda$ plane in Figure~\ref{Fig5}. We
find that a decelerating expanding universe with $\Omega_\Lambda=0$
is only accommodated by observations at $99.7\%$ C. L. when the
spherical $\beta$ model is considered. That is, a positive
cosmological constant is still needed to account for the dimming of
SNe Ia and the standard cosmological scenario remains being
supported by current observations.

Finally, it should be pointed out that, the presence of systematic
uncertainties in observations, especially ADD measurements using
SZE+X-ray surface brightness observations, and the assumption of the
frequency independency of absorption in the observed frequency range
in our analysis might result in some biases of our test results.
Moreover, as for the DD relation test, the morphological models of
galaxy cluster may also exert a remarkable influence on the tests
for cosmic opacity. In fact, any conclusion that current data may
favor a non-zero opacity should be backed up with a thorough
analysis of these systematics. Therefore, we may expect more
vigorous and convincing constraints on the cosmic opacity within the
coming years with more precise data, especially the ADD data, and a
deeper understanding for the absorption in various wavelength bands
and the intrinsic three dimensional shape of clusters of galaxies.

\section*{acknowledgments} We would like to thank A. Avgoustidis for helpful discussions.
This work was supported by the National Natural Science Foundation
of China under Grants Nos. 10935013, 11175093, 11075083 and
11222545, the Ministry of Science and Technology National Basic
Science Program (Project 973) under Grant No.2012CB821804, Zhejiang
Provincial Natural Science Foundation of China under Grants Nos.
Z6100077 and R6110518, the FANEDD under Grant No. 200922, the
National Basic Research Program of China under Grant No.
2010CB832803, the NCET under Grant No. 09-0144,  the PCSIRT under
Grant No. IRT0964, the Hunan Provincial Natural Science Foundation
of China under Grant No. 11JJ7001, and the SRFDP under Grant
No.20124306110001. ZL was partially supported by China Postdoc Grant
No .2013M530541.

\begin{table}[!h]
\begin{center}
\begin{tabular}{c}
\hline
~~~gas distribution model~~~~~~~~~~~~~~~~~~~~~~$\tau(z)=2\epsilon z$~~~~~~~~~~~~~~~~~~~~$\tau(z)=(1+z)^{2\epsilon}-1$\\
\hline
~~~~~~~~~~~Elliptical $\beta$ model~~~~~~~~~~~~~~~~~$\epsilon=0.009^{+0.059+0.127+0.199}_{-0.055-0.110-0.160}$~~~~~$\epsilon=0.014^{+0.071+0.145+0.219}_{-0.069-0.138-0.203}$\\
~~~~~~~~~~~Spherical $\beta$ model~~~~~~~~~~~~~~~~~$\epsilon=0.081^{+0.046+0.100+0.158}_{-0.042-0.085-0.124}$~~~~~$\epsilon=0.096^{+0.058+0.114+0.169}_{-0.056-0.107-0.154}$\\
Spherical $\beta$ model($r<100$ kpc-cut)~~~~~$\epsilon=0.120^{+0.047+0.101+0.143}_{-0.043-0.086-0.114}$~~~~~$\epsilon=0.140^{+0.054+0.110+0.148}_{-0.052-0.102-0.142}$\\
~~~hydrostatic equilibrium model~~~~~~~~~~$\epsilon=0.066^{+0.037+0.079+0.123}_{-0.035-0.070-0.102}$~~~~~$\epsilon=0.080^{+0.046+0.090+0.135}_{-0.045-0.086-0.126}$\\
\hline
\end{tabular}
\end{center}
\caption{\label{Tab1}Summary of the results for different optical
depth parameterizations and cluster gas distribution models. }
\end{table}


\begin{thebibliography}{99}
\bibitem{Riess0}A. G. Riess {\it et al.}, AJ, 116, 1009 (1998).
\bibitem{Perlmutter}S. Perlmutter {\it et al.}, ApJ, 517, 565 (1999).
\bibitem{Aguirre}A. Aguirre, ApJ 525, 583 (1999).
\bibitem{Csaki} C. Csaki, N. Kaloper and J. Terning, Phys. Rev. Lett. 88, 161302 (2002).
\bibitem{Tolman}R. C. Tolman, Proc. Natl. Acad. Sci., 16, 511 (1930).
\bibitem{Etherington1}I. M. H. Etherington, Phil. Mag., 15, 761 (1933).
\bibitem{Etherington2}I. M. H. Etherington, Gen. Rel. Grav., 39, 1055 (2007).
\bibitem{Ellis}G. F. R. Ellis, Gen. Rel. Grav., 39, 1047 (2007).
\bibitem{Schneider}P. Schneider, J. Ehlers and E. E. Falco, Gravitational Lenses (New York: Springer, 1999).
\bibitem{Cunha}J. V. Cunha, L. Marassi and J. A. S.  Lima, MNRAS, 379, L1 (2007).
\bibitem{Mantz}A. Mantz {\it et al.}, MNRAS, 406, 1773 (2010).
\bibitem{Komatsu}E. Komatsu {\it et al.}, ApJS, 192, 18 (2011).
\bibitem{Bassett1}B. A. Bassett and M.  Kunz, Phys. Rev. D 69, 101, 305 (2004).
\bibitem{Bassett2}B. A. Bassett and M. Kunz, ApJ, 607, 661 (2004).
\bibitem{Uzan}J. P. Uzan, N. Aghanim and Y. Mellier, Phys. Rev. D 70, 083553 (2004).
\bibitem{De bernardis}F. De Bernardis, E. Giusarma and A. Melchiorri, Int. J. Mod. Phys. D 15, 759 (2006).
\bibitem{Holanda1}R. F. L. Holanda, J. A. S. Lima and M. B. Ribeiro, ApJ, 722, L233 (2010).
\bibitem{Holanda2}R. F. L. Holanda, J. A. S. Lima and  M. B. Ribeiro, A\&A, 528, L14 (2011).
\bibitem{Zhengxiang}Z. Li, P. Wu and H. Yu, ApJ, 729, L14 (2011).
\bibitem{Nair}R. Nair, S. Jhingan, and D. Jain, JCAP05(2011)023.
\bibitem{Amanullah}R. Amanullah {\it et al.}, ApJ, 716, 712 (2010).
\bibitem{Filippis}E. De Filippis, M. Sereno, W. Bautz and G. Longo, ApJ, 625, 108 (2005).
\bibitem{Bonamente}M. Bonamente {\it et al.}, ApJ, 647, 25 (2006).
\bibitem{Hu}X. Meng, T. Zhang, H. Zhan and X. Wang, ApJ, 745, 98 (2012).
\bibitem{Liang}N. Liang, S. Cao, K. Liao and Z. Zhu, arXiv:1104.2497.
\bibitem{Cardone}V. F. Cardone, S. Spiro, I. Hook and R. Scaramella, Phys. Rev. D, 85, 123510 (2012).
\bibitem{Holandad}R. F. L. Holanda, J. A. S. Lima and M. B. Ribeiro, A\&A, 538, A131 (2011).
\bibitem{Holandac}R. F. L. Holanda, R. S. Goncalves and J. S. Alcaniz, arXiv:1201.2378.
\bibitem{Goncalves}R. S. Goncalves, R. F. L. Holanda and J. S. Alcaniz, MNRAS, 420, L43 (2012).
\bibitem{Chen1}B. Chen and R. Kantowski, Phys. Rev. D 79, 104007 (2009).
\bibitem{Chen2}B. Chen and R. Kantowski, Phys. Rev. D 80, 044019 (2009).
\bibitem{More}S. More, T. Bovy and D. W. Hogg, ApJ, 696, 1727 (2009).
\bibitem{Davis}T. M. Davis {\it et al.}, ApJ, 666, 716 (2007).
\bibitem{Percival}W. J. Percival {\it et al.}, MNRAS, 381, 1053 (2007).
\bibitem{Jun}J. Chen, P. Wu, H. Yu and Z. Li, JCAP10(2012)029.
\bibitem{Nair2}R. Nair, S. Jhingan, and D. Jain, JCAP12(2012)028.
\bibitem{Avgoustidis1}A. Avgoustidis, L. Verde and R. Jimenez, JCAP06(2009)012.
\bibitem{Avgoustidis2}A. Avgoustidis {\it et al.}, JCAP10(2010)024.
\bibitem{Kowalski}M. Kowalski {\it et al.}, ApJ, 686, 749 (2008).
\bibitem{Stern}D. Stern, {\it et al.}, JCAP02(2010)008
\bibitem{Riess1}A. G. Riess {\it et al.}, ApJ, 699, 539 (2009).
\bibitem{Lima}J. A. S. Lima, J. V. Cunha and V. T. Zanchin, ApJ, 742, L26 (2011).
\bibitem{Sunyaev}R. A. Sunyaev and Ya. B. Zel'dovich, Comments Astrophys. Space Phys., 4, 173 (1972).
\bibitem{Cavaliere}A. Cavaliere and R. Fusco-Fermiano, A\&A, 667, 70 (1978).
\bibitem{SCP}http://supernova.lbl.gov/union/.
\bibitem{Suzuki}N. Suzuki {\it et al.}, ApJ, 746, 85 (2012).
\bibitem{Mason}B. S. Mason, S. T. Myers and A. C. Readhead, ApJ, 555, L11 (2001).
\bibitem{Reese}E. D. Reese {\it et al.}, ApJ, 581, 53 (2002).
\bibitem{Bevington}P. R. Bevington and D. K. Robinson, {\it Data reduction and error analysis for the
physical sciences}, (McGraw-Hill, Boston, MA, 2003), 3rd ed.,
Chapter 4.
\end{thebibliography}
\end{document}